\documentclass[prd,onecolumn,nofootinbib,showpacs,showkeys]{revtex4}
\usepackage{bm,amsmath,amssymb}
\newcommand\gothfamily{\usefont{U}{ygoth}{m}{n}}
\DeclareTextFontCommand{\textgoth}{\gothfamily}

\begin{document}

\title{On the polarization of nonlinear gravitational waves}
\author{Nikodem J. Pop{\l}awski}
\affiliation{Department of Physics, Indiana University, Swain Hall West, 727 East Third Street, Bloomington, Indiana 47405, USA}
\email{nipoplaw@indiana.edu}
\date{\today}

\begin{abstract}
We derive a relation between the two polarization modes of a plane, linear gravitational wave in the second-order approximation.
Since these two polarizations are not independent, an initially monochromatic gravitational wave loses its periodic character due to the nonlinearity of the Einstein field equations.
Accordingly, real gravitational waves may differ from solutions of the linearized field equations, which are being assumed in gravitational-wave detectors.
\end{abstract}
\pacs{04.30.Nk}
\keywords{gravitational wave, polarization, second order, quadratic approximation.}
\maketitle

Let us consider the metric tensor whose components are functions of only one variable $u=x^0$:
\begin{equation}
g_{ik}=g_{ik}(u),
\end{equation}
where the Latin indices denote all the coordinates.
Let also assume that $\mbox{det}(g_{\alpha\beta})=0$, where the Greek indices refer to the coordinates $1,2,3$.
In this case, it can be shown (\cite{LL}, \S 109) that a suitable transformation of the coordinates:
\begin{equation}
x^0\rightarrow\phi^0(u),\,\,\,\,x^\alpha\rightarrow x^\alpha+\phi^\alpha(u),
\end{equation}
which preserves the dependence of $g_{ik}$ only on $u$, can bring the components $g_{00},g_{02},g_{03}$ to zero, and $g_{01}$ to a constant, which we set to $1/2$ \cite{Niko}.
The interval has therefore the form
\begin{equation}
ds^{2}=du\,dv+g_{ab}(u)(dx^{a}+g^a dx^1)(dx^{b}+g^b dx^1),
\label{interv}
\end{equation}
where $v=x^1$ and $g^a$ is a two-dimensional vector.
The coordinate $x^0$ has a lightlike character: $dx^\alpha=0$ and $dx^0\neq0$ yield $ds=0$.

The field equations in vacuum, $R_{ab}=0$, give $g^a=$ const, so the interval (\ref{interv}) can be brought, by a transformation $x^a+g^a x^1\rightarrow x^a$, to the form \cite{LL}
\begin{equation}
ds^{2}=du\,dv+g_{ab}(u)dx^{a}dx^{b},
\label{int}
\end{equation}
where the indices $a,b$ refer to the coordinates $2,3$.
If the interval is given by (\ref{int}) then all components of the Ricci tensor vanish identically except
\begin{equation}
R_{uu}=-\frac{1}{2}{\dot{\kappa}}^{a}_{a}-\frac{1}{4}\kappa^{a}_{b}\kappa^{b}_{a},
\label{field}
\end{equation}
where: $\kappa_{ab}={\dot{g}}_{ab}$, $\kappa^a_b=g^{ac}\kappa_{bc}$, $g^{ab}$ is the inverse of $g_{ab}$: $g^{ac}g_{bc}=\delta^a_b$, and the dot denotes differentiation with respect to $u$.
Thus, the Einstein field equations in the case of a plane (but not constrained to be weak) gravitational wave reduce to one equation, $R_{uu}=0$, for three functions $g_{22}(u),g_{23}(u),g_{33}(u)$.
A strong gravitational wave is characterized by two arbitrary functions of $u$ (\cite{LL}, \S 109).
These two functions automatically satisfy the wave equation because $u$ is a lightlike variable.
If we substitute $u=ct-x$ and $v=ct+x$ then we reproduce, in the limit of small $\kappa_{ab}$, a weak gravitational wave (\cite{LL}, \S 107) propagating in the direction of the $x$-axis.\footnote{
The number of the functionally independent field equations may seem smaller than the number of variables (the independent components of the metric tensor).
While a weak gravitational wave is described by two arbitrary functions that satisfy a wave equation resulting from the Einstein field equations in vacuum, a strong gravitational wave is described by two arbitrary functions that satisfy a wave equation resulting from the assumption $g_{ik}=g_{ik}(x^0)$.
}

Let the two-dimensional metric tensor $g_{ab}$ of a plane gravitational wave be given by
\begin{equation}
g_{ab}= \left[ \begin{array}{cc}
-1+f(u) & h(u) \\
h(u) & -1+g(u) \end{array} \right],
\label{metric}
\end{equation}
where $f,g,h$ are small quantities \cite{Niko}.
We therefore have
\begin{equation}
\kappa_{ab}= \left[ \begin{array}{cc}
\dot{f} & \dot{h} \\
\dot{h} & \dot{g} \end{array} \right].
\end{equation}
To calculate $\kappa^a_b$ in the quadratic approximation (keeping only linear and quadratic terms in $f,g,h$), we only need the linear approximation in $g^{ab}$ because $\kappa_{ab}$ is already on the order of $f,g,h$:
\begin{equation}
g^{ab}= \left[ \begin{array}{cc}
-1-f & -h \\
-h & -1-g \end{array} \right].
\end{equation}
We obtain
\begin{equation}
\kappa^a_b= \left[ \begin{array}{cc}
-(1+f)\dot{f}-h\dot{h} & -h\dot{f}-(1+g)\dot{h} \\
-h\dot{g}-(1+f)\dot{h} & -(1+g)\dot{g}-h\dot{h} \end{array} \right].
\end{equation}
Equation (\ref{field}) in the quadratic approximation becomes
\begin{equation}
R_{uu}=\frac{1}{4}\bigl(2(1+f)\ddot{f}+2(1+g)\ddot{g}+\dot{f}^2+\dot{g}^2+4h\ddot{h}+2\dot{h}^2\bigr),
\end{equation}
so in vacuum we have \cite{Niko}
\begin{equation}
2(1+f)\ddot{f}+2(1+g)\ddot{g}+\dot{f}^2+\dot{g}^2+4h\ddot{h}+2\dot{h}^2=0.
\label{FE}
\end{equation}

In the linear approximation, the field equation (\ref{FE}) reads $\ddot{f}+\ddot{g}=0$, from which it follows that $f+g$ is a linear function of $u$.
If both $f$ and $g$ are periodic functions of $u$ then this condition becomes $f+g=0$.
Denoting $f$ by $h_+$ and $h$ by $h_\times$ turns the metric tensor (\ref{metric}) into
\begin{equation}
g_{ab}= \left[ \begin{array}{cc}
-1+h_+ & h_\times \\
h_\times & -1-h_+ \end{array} \right].
\label{polar}
\end{equation}
The quantities $h_+$ and $h_\times$ are two independent polarizations and are given by
\begin{equation}
h_+=-\frac{G}{3c^4R}(\ddot{D}_{22}-\ddot{D}_{33}),\,\,\,\,h_\times=-\frac{2G}{3c^4R}\ddot{D}_{23},
\end{equation}
where $D_{ab}$ are the components of the quadrupole moment of a gravitationally radiating source and $R$ is the distance from such a source (\cite{LL}, \S 110).

In the quadratic approximation, the field equation (\ref{FE}) does not yield the condition $f+g=0$.
Instead, it leads to a more complicated relation between $f$, $g$ and $h$.
If we impose the condition $f+g=0$ (which is being assumed in current gravitational-wave-signal analyses) in order to define $h_+=f$ and $h_\times=h$, then (\ref{FE}) becomes
\begin{equation}
2f\ddot{f}+\dot{f}^2+2h\ddot{h}+\dot{h}^2=0,
\end{equation}
or
\begin{equation}
2h_+\ddot{h_+}+\dot{h_+}^2+2h_\times\ddot{h_\times}+\dot{h_\times}^2=0.
\label{consist}
\end{equation}
Equation (\ref{consist}) is the consistency condition that a plane gravitational wave, provided that $f+g=0$, satisfies in the quadratic approximation.

In the current search for gravitational waves, it is assumed that a gravitational wave consist of two independent modes of polarization in the transverse-transpose gauge, $h_{+}$ and $h_{\times}$, and the detector response is given by a linear combination \cite{review}
\begin{equation}
\xi(t)=F_{+}h_{+}(t)+F_{\times}h_{\times}(t).
\end{equation}
This assumption of two independent polarizations is true only in the linear approximation.
At higher orders, the two polarizations are no longer independent and must obey (at the second order) the consistency condition (\ref{consist}).
This consistency condition should be included in all gravitational-wave-signal analyses.
It could also enhance the detection efficiency in burst searches that look for gravitational waves from transient sources without assuming the form of a signal.
Some examples of burst sources include core-collapse supernovae, gamma-ray bursts and black hole mergers \cite{bursts}.

If we assume that, at $u=0$, we have a monochromatic plane gravitational wave in the $h_+$ polarization ($h_\times=0$): $f=A\mbox{cos}(ku)$, where $k=2\pi/\lambda$ and $\lambda$ is the wavelength, then (\ref{consist}) reads
\begin{equation}
2h_\times\ddot{h_\times}+\dot{h_\times}^2=A^2k^2(3\mbox{cos}^2(ku)-1).
\label{grow}
\end{equation}
As a result, $h_\times$ becomes different from zero and such a wave becomes a combination of the two polarizations.
Moreover, (\ref{grow}) gives a non-periodic solution for $h_\times(u)$, so the initial $h_+$ gravitational wave loses its periodic-wave character.
A similar analysis shows that an initial $h_\times$ wave exhibits the same behavior.
The lost of a periodic character due to quadratic corrections happens after the wave, initially monochromatic and polarized (or in a combination of the two polarizations), propagates through distances on the order of $\lambda$.

As a final remark, we note that the necessity of using the second-order corrections in a consistent treatment of gravitational waves and their propagation of energy and momentum has been discussed in \cite{nonlinear}.

\begin{acknowledgements}
The author would like to thank Shantanu Desai for fruitful discussions on gravitational waves.
\end{acknowledgements}


\begin{thebibliography}{}
\bibitem{LL} L. D. Landau and E. M. Lifshitz, {\em The Classical Theory of Fields} (Pergamon, New York, 1975).
\bibitem{Niko} N. J. Pop{\l}awski, J. Math. Phys. {\bf 47}, 072501 (2006).
\bibitem{review} E. E. Flanagan and S. A. Hughes, New J. Phys. {\bf 7}, 204 (2005); A. Buonanno, arXiv:0709.4682.
\bibitem{bursts} C. Cutler and K. S. Thorne, arXiv:gr-qc/0204090.
\bibitem{nonlinear} R. Aldrovandi, J. G. Pereira, R. da Rocha, and K. H. Vu, Int. J. Theor. Phys. {\bf 49}, 549 (2010).
\end{thebibliography}
\end{document}